\documentclass[9pt,twocolumn,twoside]{gsajnl}

\usepackage[utf8]{inputenc}
\usepackage[T1]{fontenc}
\usepackage{amsfonts}
\usepackage{amsmath}
\usepackage[super]{natbib}
\usepackage{hyperref}
\hypersetup{
    colorlinks=true,
    linkcolor=blue,
    filecolor=magenta,      
    urlcolor=cyan,
    pdftitle={Overleaf Example},
    pdfpagemode=FullScreen,
    }
\usepackage[capitalise]{cleveref}
\usepackage{placeins}

\usepackage{graphicx}
\usepackage{booktabs}

\usepackage{caption}        % Subfigure packages
\usepackage{subcaption}     % Subfigure packages
\usepackage{multirow}
\usepackage{xcolor}

\usepackage{microtype}
\usepackage{contour}
\usepackage{adjustbox}

\newcommand{\absdiv}[1]{%
  \par\addvspace{.5\baselineskip}% adjust to suit
  \noindent\textbf{#1}\quad\ignorespaces
}

\renewenvironment{thebibliography}[1]{
  \begin{oldthebibliography}{#1}
    \setlength{\itemsep}{0em}
    \setlength{\parskip}{0em}
}
{
  \end{oldthebibliography}
}

\title{An open-source deep learning algorithm for efficient and fully-automatic analysis of the choroid in optical coherence tomography}
\runningtitle{Fully-automatic choroid segmentation in OCT}
\runningauthor{Burke et al.}

\author[$\ast$,1]{Jamie Burke}
\author[$\ast$,2,3]{Justin Engelmann}
\author[4]{Charlene Hamid}
\author[4]{Megan Reid-Schachter}
\author[5]{Tom Pearson}
\author[6]{Dan Pugh}
\author[6]{Neeraj Dhaun}
\author[7]{Amos Storkey}
\author[1]{Stuart King}
\author[4,8]{Tom MacGillivray}
\author[3,9]{Miguel O. Bernabeu}
\author[10,7]{Ian J.C. MacCormick}

\affil[1]{School of Mathematics, University of Edinburgh, Edinburgh, UK}
\affil[2]{School of Informatics, University of Edinburgh, Edinburgh, UK}
\affil[3]{Centre for Medical Informatics, University of Edinburgh, Edinburgh, UK}
\affil[4]{Clinical Research Facility and Imaging, University of Edinburgh, Edinburgh, UK}
\affil[5]{University Hospital Wales, NHS Wales, Cardiff, Wales, UK}
\affil[6]{British Heart Foundation Centre for Cardiovascular Science, University of Edinburgh, Edinburgh, UK}
\affil[7]{Institute for Adaptive and Neural Computation, School of Informatics, University of Edinburgh, Edinburgh, UK}
\affil[8]{Centre for Clinical Brain Sciences, University of Edinburgh, Edinburgh, UK }
\affil[9]{The Bayes Centre, University of Edinburgh, Edinburgh, UK}
\affil[10]{Centre for Inflammation Research, The Queen’s Medical Research Institute, University of Edinburgh, Edinburgh, UK}

\correspondingauthoraffiliation[$\ast$]{Corresponding and lead authors with equal contribution; \\ Email address: \url{Jamie.Burke@ed.ac.uk} \& \url{Justin.Engelmann@ed.ac.uk}}

\begin{abstract}
\absdiv{Purpose} 
To develop an open-source, fully-automatic deep learning algorithm, DeepGPET, for choroid region segmentation in optical coherence tomography (OCT) data.  

\absdiv{Methods}
We used a dataset of 715 OCT B-scans (82 subjects, 115 eyes) from 3 clinical studies related to systemic disease. Ground truth segmentations were generated using a clinically validated, semi-automatic choroid segmentation method, Gaussian Process Edge Tracing (GPET). We finetuned a UNet with MobileNetV3 backbone pre-trained on ImageNet. Standard segmentation agreement metrics, as well as derived measures of choroidal thickness and area, were used to evaluate DeepGPET, alongside qualitative evaluation from a clinical ophthalmologist. 
 
\absdiv{Results} 
DeepGPET achieves excellent agreement with GPET on data from 3 clinical studies (AUC=0.9994, Dice=0.9664; Pearson correlation of 0.8908 for choroidal thickness and 0.9082 for choroidal area), while reducing the mean processing time per image on a standard laptop CPU from 34.49s (±15.09) using GPET to 1.25s (±0.10) using DeepGPET. Both methods performed similarly according to a clinical ophthalmologist, who qualitatively judged a subset of segmentations by GPET and DeepGPET, based on smoothness and accuracy of segmentations.

\absdiv{Conclusions}
DeepGPET, a fully-automatic, open-source algorithm for choroidal segmentation, will enable researchers to efficiently extract choroidal measurements, even for large datasets. As no manual interventions are required, DeepGPET is less subjective than semi-automatic methods and could be deployed in clinical practice without necessitating a trained operator.

\end{abstract}

\begin{document}

\maketitle     

\section{Introduction}
The retinal choroid is a complex, extensively interconnected vessel network positioned between the retina and the sclera. The choroid holds the majority of the vasculature in the eye and plays a pivotal role in nourishing the retina. Optical coherence tomography (OCT) is an ocular imaging modality that uses low-coherence light to construct a three-dimensional map of chorioretinal structures at the back of the eye. Standard OCT imaging does not visualise the deeper choroidal tissue well as it sits beneath the hyperreflective retinal pigment epithelium layer of the retina. Enhanced Depth Imaging OCT (EDI-OCT) overcomes this problem and offers improved visualisation of the choroid, thus providing a unique window into the microvascular network which not only resides closest to the brain embryologically, but also carries the highest volumetric flow per unit tissue weight compared to any other organ in the body. 

Since the advent of OCT, interest in the role played by the choroid in systemic health has been growing \cite{tan2016state}, as non-invasive imaging of the choroidal microvasculature may provide a novel location to detect systemic, microvascular changes early. Indeed, changes in choroidal blood flow, thickness and other markers have been shown to correspond with patient health such as choroidal thickness in chronic kidney disease \cite{balmforth2016chorioretinal} and choroidal area and vascularity in Alzheimer's dementia \cite{robbins2021choroidal}.

Quantification of the choroid in EDI-OCT imaging requires segmentation of the choroidal space. However, this is a harder problem than retinal layer segmentation due to poor signal penetration from the device --- and thus lower signal-to-noise ratio --- and shadows cast by superficial retinal vessels and choroidal stroma tissue. This results in poor intra- and inter-rater agreement even with manual segmentation by experienced clinicians, and manual segmentation is too labour intensive and subjective to be practical for analysing large scale datasets. 

Semi-automated algorithms improve on this slightly but are typically multi-stage procedures, requiring traditional image processing techniques to prepare the images for downstream segmentation \cite{eghtedar2022update}. Methods based on graph theory such as Dijkstra's algorithm \cite{masood2018automatic,salafian2018automatic} or graph cut \cite{kajic2012automated}, as well as on statistical techniques including level sets \cite{wang2017automatic,srinath2014automated}, contour evolution \cite{george2019two}, and Gaussian mixture models \cite{danesh2014segmentation} have been proposed previously. Concurrently, deep learning(DL)-based approaches have emerged. \cite{chen2017automated} used a DL model for choroid layer segmentation, but with traditional contour tracing as a post-processing step. Other DL-based approaches, too, combine traditional image processing techniques as pre- or post-processing steps \cite{sui2017choroid,masood2019automatic,al2017novel} whereas others are fully DL-based \cite{chen2022application,zheng2021deep}, the latter of which is in a similar vein to the proposed method. More recently, DL has been used to distil existing semi-automatic traditional image processing pipelines into a fully-automatic method \citep{engelmann2022robust}.

Gaussian Process Edge Tracing (GPET), based on Bayesian machine learning \cite{burke2021edge}, is a particularly promising method for choroid layer segmentation that has been clinically and quantitatively validated \cite{burke2023evaluation}. Gaussian process (GP) regression is used to model the upper and lower boundaries of the choroid from OCT scans. For each boundary, a recursive Bayesian scheme is employed to iteratively detect boundary pixels based on the image gradient and the GP regressor's distribution of candidate boundaries. However, GPET is semi-automatic and thus requires time-consuming manual interventions by specifically trained personnel which introduces subjectivity and limits the potential for analysing larger datasets or deploying GPET into clinical practice. 

There are currently no accessible, open-source algorithms for fully-automatic choroidal segmentation. 
All available algorithms fall into one of three categories: First, semi-automatic methods \cite{ss2019octtools, brandt2018octmarker} that require human supervision and thus require training and introduce subjectivity. Second, fully-automatic DL-based methods that are not openly accessible, either only providing the code but not the trained model necessary to use the method \cite{kugelman2019automatic} or not providing any access at the time of writing \cite{xuan2023deep}. Third, fully-automatic but comprising of many steps, requiring a good understanding of image processing techniques and a license for proprietary software (MATLAB) \cite{mazzaferri2017open}. 

\begin{figure*}[!tb]
    \centering
    \includegraphics[width=\textwidth]{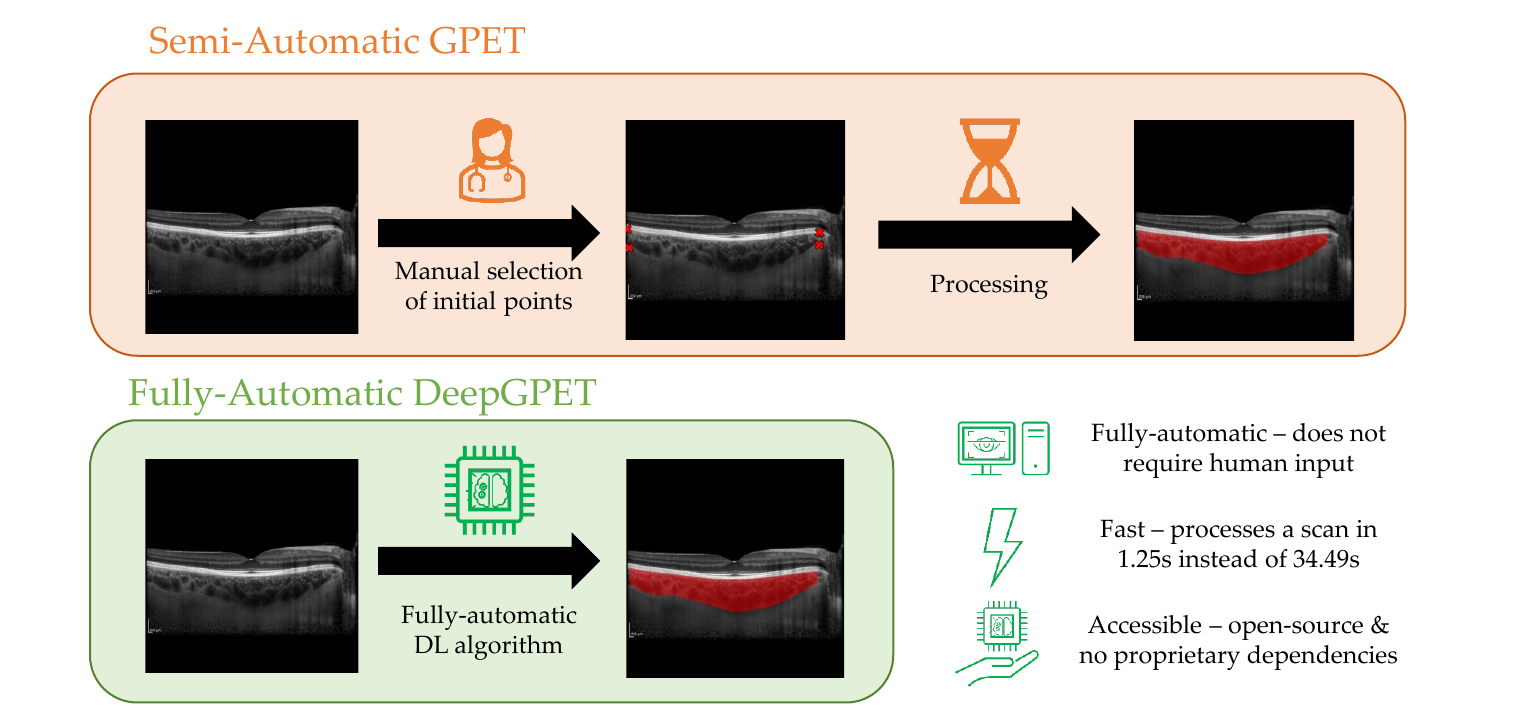}
    \caption{Comparison between the semi-automatic GPET \cite{burke2021edge, burke2023evaluation} (top) and fully-automatic DeepGPET (bottom).}
    \label{fig:info_fig}
\end{figure*}

We aim to develop and release an open-source, raw image-to-measurement, fully-automatic method for choroid region segmentation that can be easily used without special training and does not require licenses for proprietary software (\cref{fig:info_fig}). Importantly, we intend to not only to make our method available to the research community, but to do so in a frictionless way that allows other researchers to download and use our method without seeking our approval. We distil GPET into a deep learning algorithm, DeepGPET, which can process images without supervision in a fraction of the time --- permitting analysis of large scale datasets and potential deployment into clinical care and research practice without prior training in image processing. The code and model weights for DeepGPET are available here: \url{https://github.com/jaburke166/deepgpet}.

\section{Methods}

\subsection{Study population}
We used 715 OCT B-scans belonging to 82 subjects from three studies: \textbf{OCTANE} \cite{dhaun2014optical}, a study looking at renal function and impairment in chronic kidney disease patients. \textbf{i-Test}, a study recruiting pregnant women of any gestation or those who have delivered a baby within 6 months, including controls and individuals at high risk of complications. \textbf{Normative}, data from 30 healthy volunteers as a control group \cite{pearson2022multi}. All studies conformed with the Declaration of Helsinki and received relevant ethical approval and informed consent from all subjects. \cref{tab:demo_tab} provides an overview of basic population characteristics and number of subjects/images of these studies. Supplementary \cref{fig:bp_chor_charac} presents box-plot distributions of choroidal thickness and area for the three datasets used to build DeepGPET, with \cref{tab:chor_characs} presenting tabular mean and standard deviation values.

Two Heidelberg spectral domain OCT SPECTRALIS devices were used for image acquisition: the Standard Module (OCT1 system) and FLEX Module (OCT2 system). The FLEX is a portable version that enables imaging of patients in a ward environment. Both machines imaged a $30^{\circ}$ (8.7 mm) region, generating a macular, cross-sectional OCT B-scan at $768\times768$ pixel resolution. Notably, 14\% of the OCT B-scans were non-EDI and thus present more challenging images with lower signal-to-noise ratio in the choroidal part of the OCT. Horizontal line and vertical scans were centred at the fovea with active eye tracking, using an Automatic Real Time (ART) value of 100. Posterior pole macular scans covered a 30-degree by 25-degree region, using EDI mode.

We split the data into approximately an 85:8:7 split between training (603 B-scans, 66 subjects), validation (58 B-scans, 9 subjects) and test sets (54 B-scans, 7 subjects). When splitting the data, we did so at the patient-level, i.e. each subjects OCT images are present in only one set, and were selected so that each set had proportionally equal amounts of scan types (EDI/non-EDI) to best represent image quality. See supplementary \cref{supp_tab:tvt_demo_tab} for an overview of basic population and imaging characteristics for each set.

\begin{table*}[tb]
\centering
{\small
\begin{tabular}{@{}lccc|c@{}}
\toprule
\hspace{3em}                    & \hspace{1em}OCTANE\hspace{1.3em}       & \hspace{1em}i-Test\hspace{1.3em}     &\hspace{1em}Normative\hspace{1.3em}  & Total        \\ \midrule
Subjects            & 47           & 5          & 30         & 82           \\
Male/Female         & 24 / 23     & 0 / 5      & 20 / 10    & 44 / 38     \\
Right/Left eyes     & 47 / 0       & 5 / 5      & 29 / 29    & 81 / 34      \\
Age (mean (SD))     & 48.8 (12.9) & 34.4 (3.4) & 49.1 (7.0) & 48.0 (11.2) \\
Machine             & Standard            & FLEX          & Standard          &    Both          \\
Horizontal/Vertical scans & 166 / 0      & 16 / 16    & 57 / 54    & 239 / 70     \\
Volume scans           & 174          & 186        & 46         & 406          \\
Total B-scans         & 340          & 218        & 157        & 715          \\ \bottomrule
\end{tabular}%
}
\caption{Overview of population characteristics. EDI, enhanced depth imaging; SD, standard deviation.}
\label{tab:demo_tab}
\end{table*}

\subsection{DeepGPET}
As the ground truths are based on GPET, DeepGPET can be can be seen as a more efficient, fully automatic and distilled version of GPET. Our approach was to fine-tune a UNet \citep{ronneberger2015u} with MobileNetV3 \citep{howard2019searching} backbone pre-trained on ImageNet for 60 epochs with batch size 16 using AdamW \citep{loshchilov2017decoupled} ($\text{lr}=10^{-3}$, $\beta_1 = 0.9$, $\beta_2 = 0.999$, $\text{weight decay}=10^{-2}$). After epoch 30, we maintain an exponential moving average (EMA) of model weights which we then use as our final model. We use the following data augmentations: brightness and contrast changes, horizontal flipping, and simulated OCT speckle noise by applying Gaussian noise followed by multiplicative noise (all $p=0.5$); Gaussian blur and random affine transforms (both $p=0.25$). To reduce memory-load, we crop the black space above and below the OCT B-scan and process images at a resolution of $544 \times 768$ pixels. Images are standardised by subtracting 0.1 and dividing by 0.2, and no further pre-processing is done. We used Python 3.11, PyTorch 2.0, Segmentation Models PyTorch \citep{Iakubovskii2019} and the timm library \citep{rw2019timm}.

\subsection{Statistical analysis}
We used Dice coefficient and Area Under the ROC Curve (AUC) for evaluating agreement in segmentations, as well as the Pearson correlation $r$ and Mean Absolute Error (MAE) for segmentation-derived choroid thickness and area. The calculation of thickness and area from the segmentation is described in more detail in \cite{burke2023evaluation}. Briefly, for thickness the average of 3 measures is used, taken at the fovea and 2,000 microns from it in either direction by drawing a perpendicular line from the upper boundary to the lower boundary to account for choroidal curvature. For area, pixels are counted in a region of interest 3,000 microns around the fovea, which corresponds to the commonly used Early Treatment Diabetic Retinopathy Study (ETDRS) macular area of $6,000 \times 6,000$ microns \cite{early1991early}.

We compare DeepGPET's agreement with GPET's segmentations against the repeatability of GPET itself. The creator of GPET, J.B., made both the original and repeated segmentations with GPET. Since both segmentations were done by the same person there is no inter-rater subjectivity at play here. Thus, the intra-rater agreement measured here is a best case scenario and forms an upper-bound for agreement with the original segmentations and any other semi-automatic method requiring manual input, which can necessarily be subject to human variability, unlike DeepGPET.

\begin{table*}[tb]
\centering
\begin{adjustbox}{max width=\textwidth}
{\small \begin{tabular}{@{}llllllll@{}}
\toprule
\multicolumn{1}{c}{\multirow{2}{*}{Method}} &
  \multicolumn{1}{c}{\multirow{2}{*}{AUC}} &
  \multicolumn{1}{c}{\multirow{2}{*}{Dice}} &
  \multicolumn{1}{c}{\multirow{2}{*}{\begin{tabular}[c]{@{}c@{}}Time \\ (s/img)\end{tabular}}} &
  \multicolumn{2}{c}{Thickness} &
  \multicolumn{2}{c}{Area} \\ \cmidrule(l){5-6}\cmidrule(l){7-8} 
\multicolumn{1}{c}{} &
  \multicolumn{1}{c}{} &
  \multicolumn{1}{c}{} &
  \multicolumn{1}{c}{} &
  \multicolumn{1}{c}{Pearson $r$} &
  \multicolumn{1}{c}{MAE ($\mu$m)} &
  \multicolumn{1}{c}{Pearson $r$} &
  \multicolumn{1}{c}{MAE (mm$^2$)} \\ \midrule
DeepGPET &\hspace{0.5em} 0.9994\hspace{0.5em} & \hspace{0.5em}0.9664\hspace{0.5em} & \hspace{0.85em}1.25 ± 0.10\hspace{0.5em}   & 0.8908 & 13.3086 & 0.9082 & 0.0699 \\ \midrule
Repeat GPET    & \hspace{0.8em}0.9812\hspace{0.5em} & \hspace{0.5em}0.9672\hspace{0.5em} & \hspace{0.5em}34.49 ± 15.09\hspace{0.5em} & 0.9527 & 10.4074 & 0.9726 & 0.0486 \\ \bottomrule
\end{tabular}%

}
\end{adjustbox}
\caption{Metrics for DeepGPET and repeated GPET using the initial GPET annotation as ``ground-truth''. Time given as mean ± standard deviation.}
%}
\label{tab:main_results}
\end{table*}

\begin{figure*}[tb]
\centering
  \begin{subfigure}{0.37\textwidth}
  \centering
  \includegraphics[width=\textwidth]{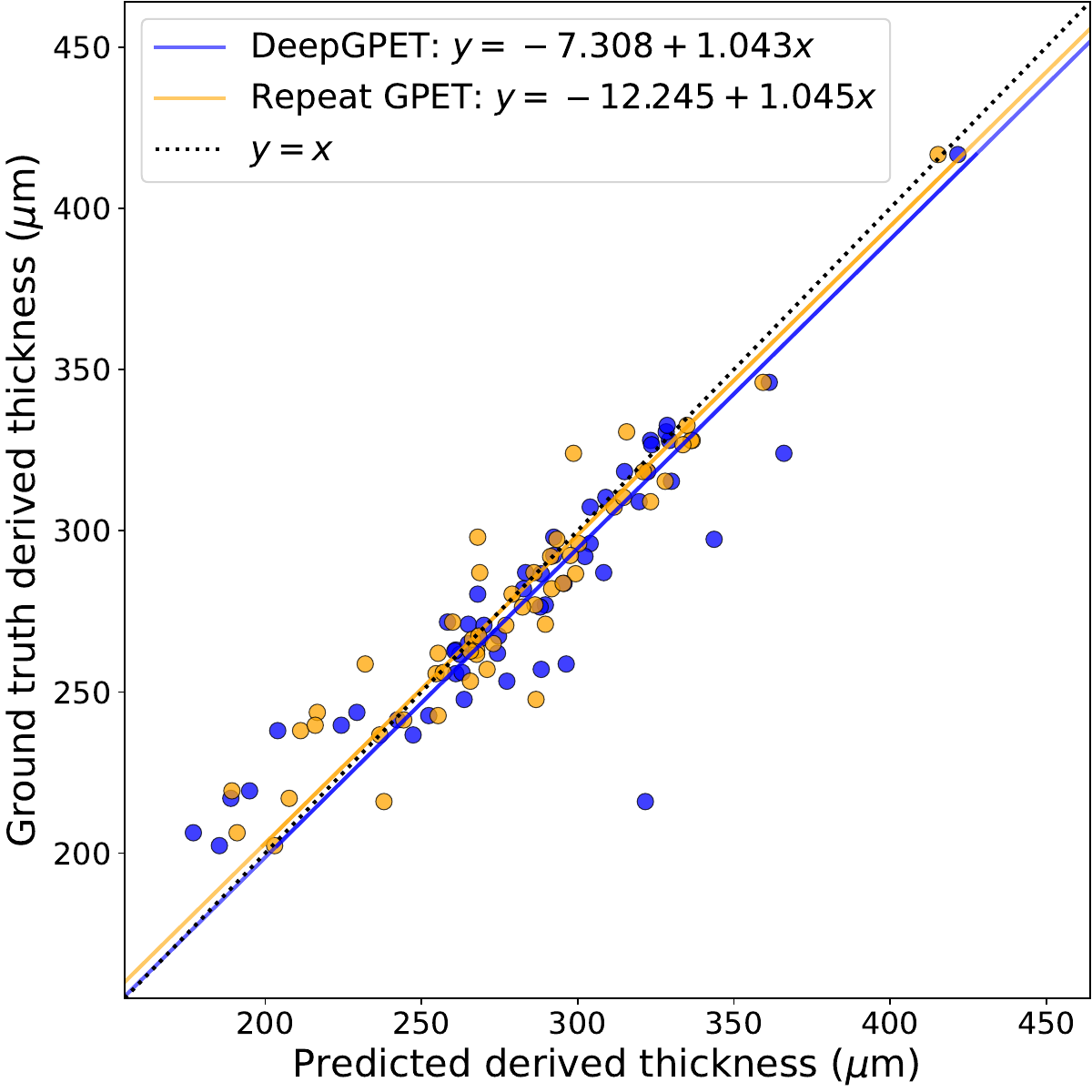}
  \end{subfigure}%
  \begin{subfigure}{0.37\textwidth}
  \centering
  \includegraphics[width=\textwidth]{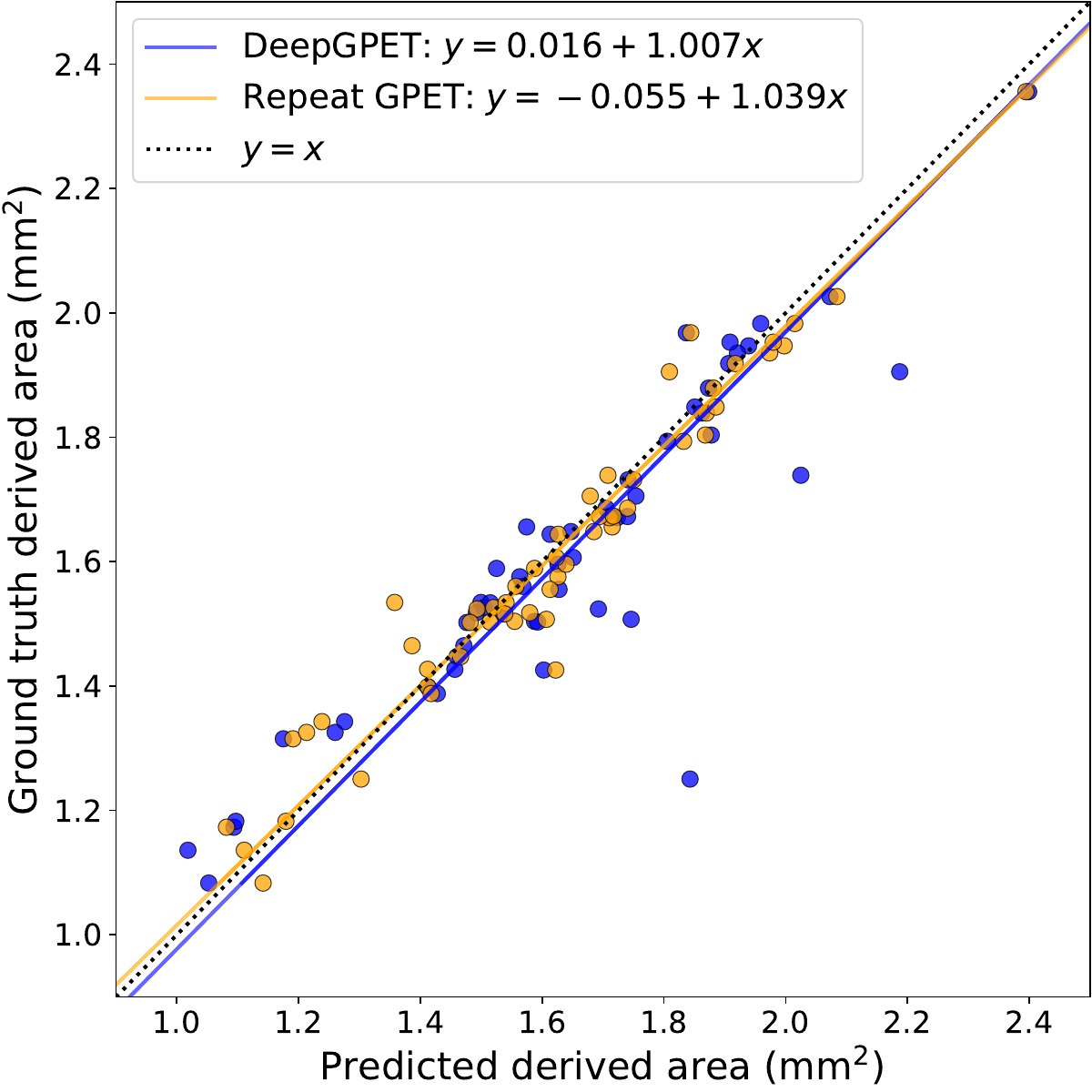}
  \end{subfigure}
\caption{Correlation plots comparing derived measures of mean choroid thickness (a) and choroid area (b) using DeepGPET and the re-segmentations using GPET.}
\label{fig:corr_plot}
\end{figure*}

In addition to quantitative evaluations, we also compared segmentations by GPET and DeepGPET for 20 test set OCT images qualitatively by having them rated by I.M., an experienced clinical ophthalmologist. We selected 7 examples with the highest disagreement in thickness and area, 7 examples with disagreement closest to the median, and 6 examples with the lowest disagreement. Thus, these 20 examples cover cases where both methods are very different, cases of typical disagreement, and cases where both methods are very similar. In each instance, I.M. was shown the segmentations of both methods overlaid on the OCT --- blinded to which method produced which segmentation --- and also provided with the raw, full-resolution OCT, and was then asked to rate each one along three dimensions: Quality of the upper boundary, the lower boundary and overall smoothness using an ordinal scale: ``Very bad'', ``Bad'', ``Okay'', ``Good'', ``Very good''. 

\section{Results}
\subsection{Quantitative}

\begin{figure*}[tb]
\centering
  \begin{subfigure}{0.37\textwidth}
  \centering
  \includegraphics[width=\textwidth]{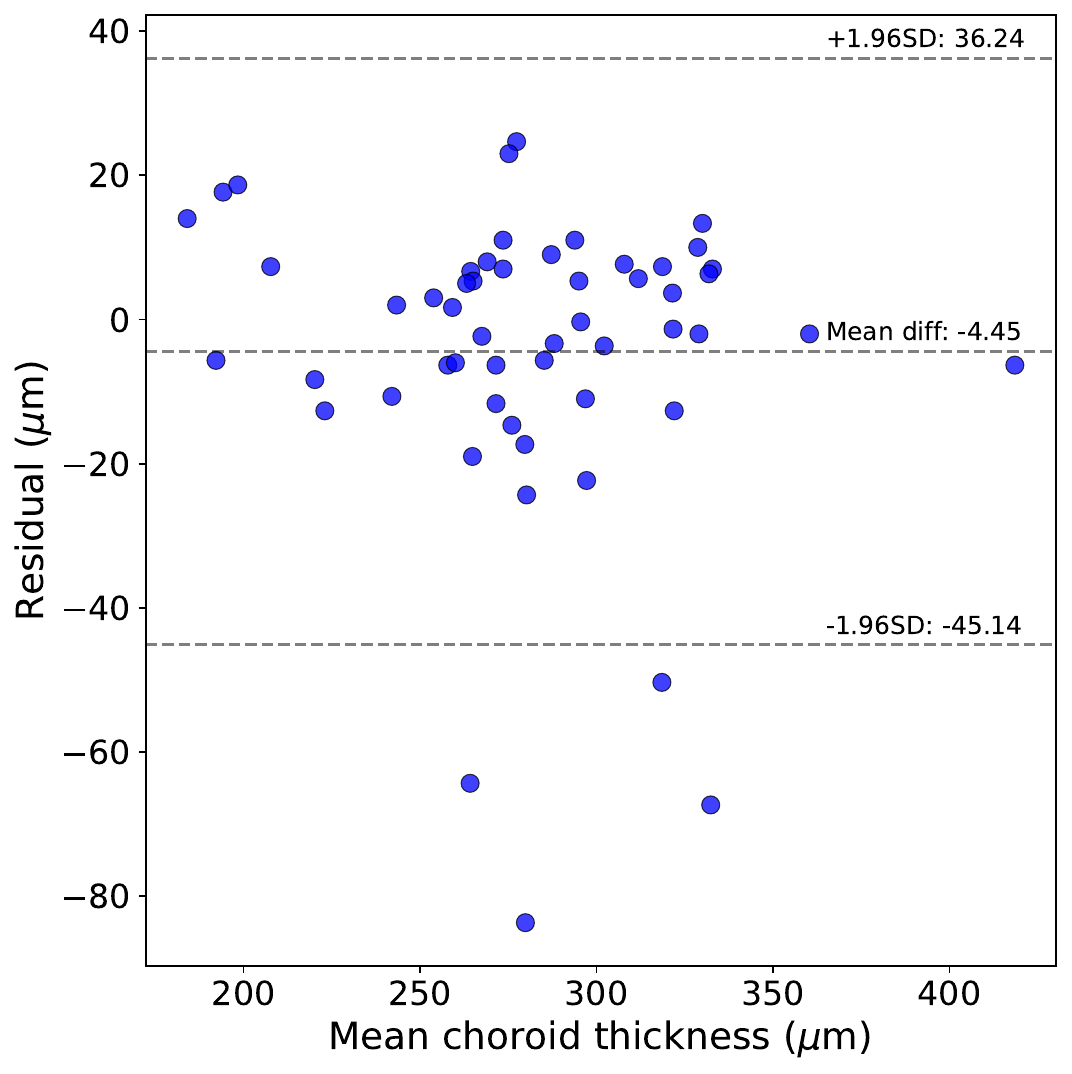}
  \end{subfigure}%
  \begin{subfigure}{0.37\textwidth}
  \centering
  \includegraphics[width=\textwidth]{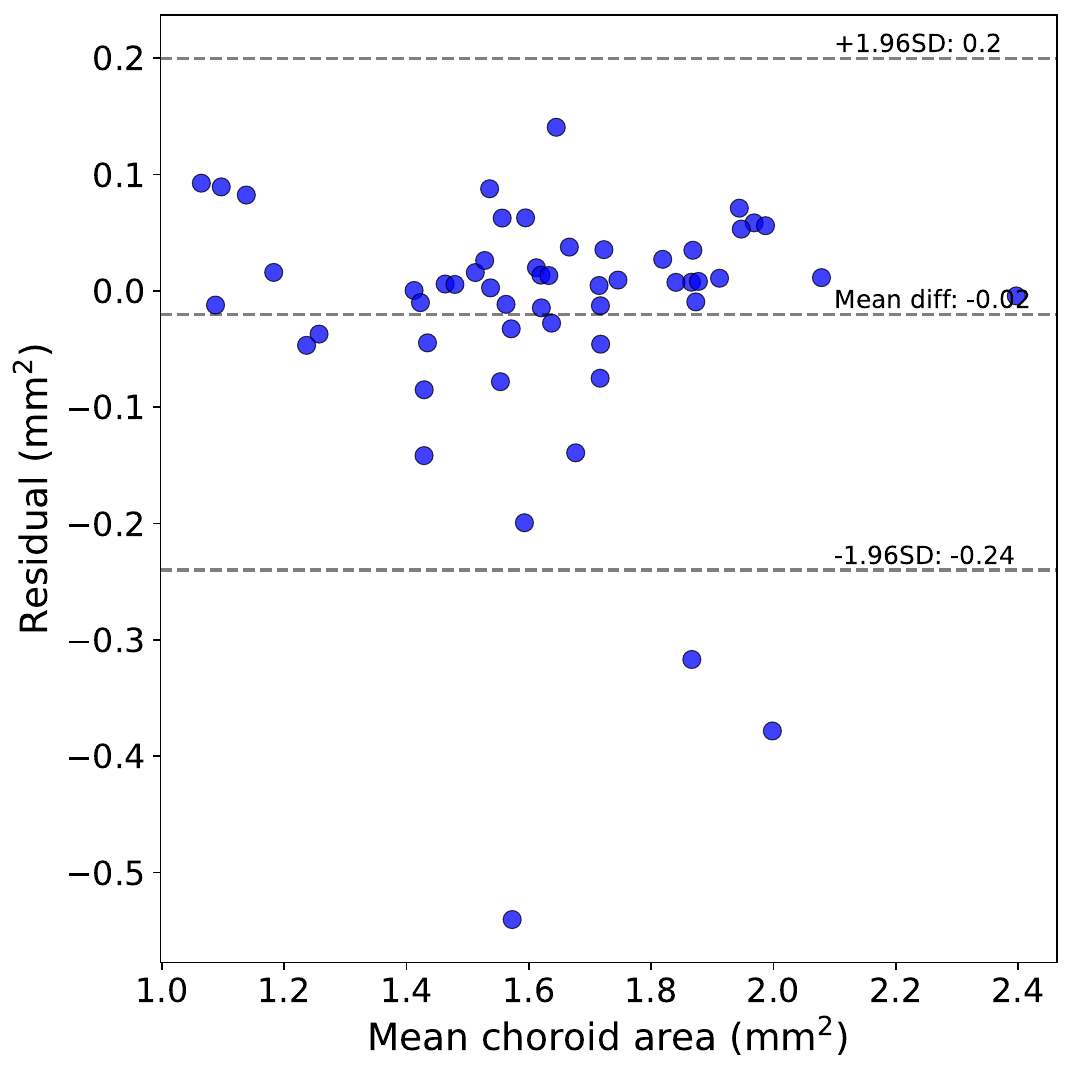}
  \end{subfigure}
\caption{Bland-altman plots comparing the agreement between DeepGPET and GPET using mean choroid thickness (a) and choroid area (b).}
\label{fig:ba_plot}
\end{figure*}

\cref{tab:main_results} shows the results for DeepGPET and a repeat GPET, compared to the initial GPET segmentation as ``ground-truth''.

\subsubsection{Agreement in segmentation.} Both methods have excellent agreement with the original segmentations. DeepGPET's agreement is comparable to the repeatability of GPET itself, with DeepGPET's AUC being slightly higher (0.9994 vs 0.9812) and Dice coefficient slightly lower (0.9664 vs 0.9672). DeepGPET performing better in terms of AUC but worse in terms of Dice suggests that for pixels where it disagrees with GPET after thresholding, the confidence is lower than for ones where it agrees with GPET. This in turn suggests that DeepGPET is well-calibrated based on the raw predictions made for each pixel.

\subsubsection{Processing speed and manual interventions.} Both methods were compared on the same standard laptop CPU. DeepGPET processed the images in only 3.6\% of the time that GPET needed. DeepGPET ran fully-automatic and successfully segmented all images, whereas GPET required 1.27 manual interventions on average, including selecting initial pixels and manual adjustment of GPET parameters when the initial segmentation failed. 

This results in massive time savings: A standard OCT volume scan consists of 61 B-scans. With GPET, processing such a volume for a single eye takes about 35 minutes during which a person has to select initial pixels to guide tracing (for all images) and adjust parameters if GPET initially failed (for about 25\% of images). In contrast, DeepGPET could do the same processing in about 76 seconds on the same hardware, during which no manual input is needed. DeepGPET could even be GPU-accelerated to cut the processing time by another order of magnitude. 

The lack of manual interventions required by DeepGPET means that no subjectivity is introduced unlike GPET, particularly when used by different people. Additionally, DeepGPET does not require specifically trained analysts and could be used fully-automatically in clinical practice.

\begin{table*}[tb]
\centering
\begin{adjustbox}{max width=\textwidth}
{\small \begin{tabular}{@{}lccc@{}}
\toprule
Method   & Upper boundary & Lower boundary                          & Smoothness                              \\ \midrule
DeepGPET & \hspace{0.5em}Very good: 20  \hspace{0.5em}& \hspace{0.5em}Very good: 4, Good: 10, Okay: 4, Bad: 2\hspace{0.5em} & \hspace{0.5em}Very good: 5, Good: 12, Okay: 2, Bad: 1 \\
GPET     & \hspace{0.5em}Very good: 20  \hspace{0.5em}& \hspace{0.5em}Very good: 6, Good: 6, Okay: 8, Bad: 0\hspace{0.5em}  & \hspace{0.5em}Very good: 6, Good: 13, Okay: 1, Bad: 0 \\ \bottomrule
\end{tabular}%
}
\end{adjustbox}
\caption{Qualitative ratings of 20 test set segmentations along 3 key dimensions. The rater was blinded to the identity of the methods and their order was randomised for every example.}
%}
\label{tab:qualitative}
\end{table*}

\begin{figure*}[tb]
    \centering
    \includegraphics[width=\textwidth]{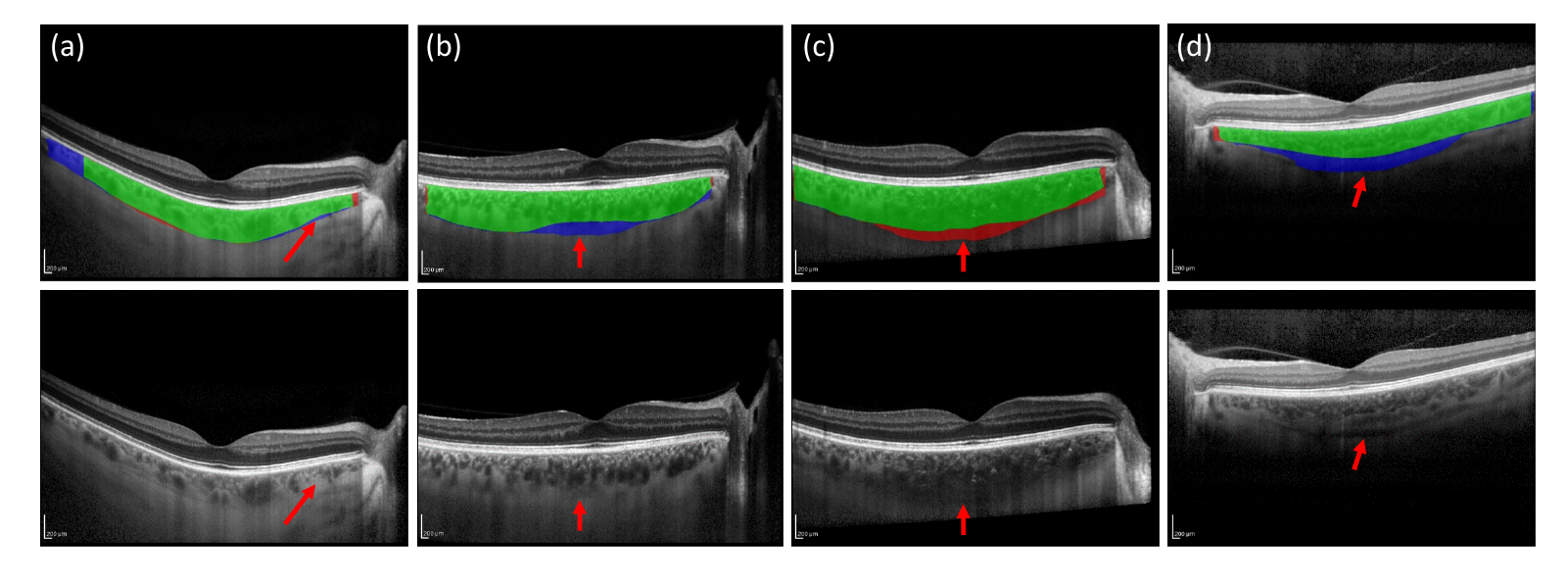}
    \caption{Four examples from the adjudication. The rater preferred DeepGPET for (a--b) and GPET for (c--d). Top row: green, segmented by both GPET and DeepGPET; red, GPET only; and blue, DeepGPET only. Bottom row: arrows indicate important choroidal features which can make segmentation challenging. (a): no large vessels in nasal region to guide segmentation; (b): lower boundary very faint and below the posterior most vessels; (c): lower boundary noisy and faint; (d): large suprachoroidal space visible.}
    \label{fig:adjudication_examples}
\end{figure*}

\begin{figure*}[!tb]
    \centering
    \includegraphics[width=\textwidth]{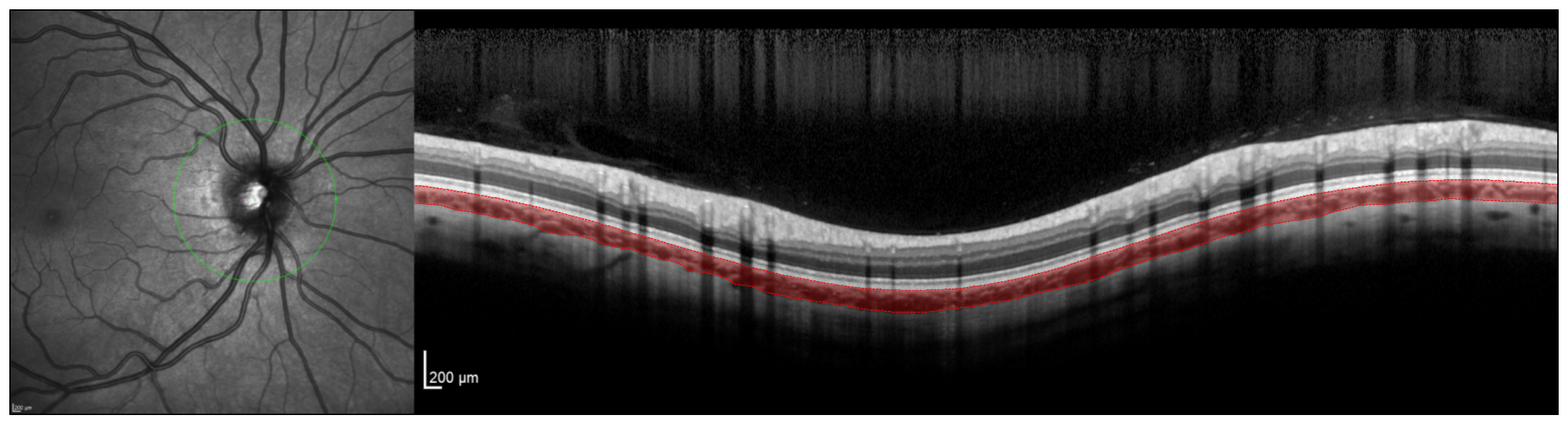}
    \caption{An example peripapillary scan from Heidelberg's Standard Module, automatically segmented by DeepGPET without manual intervention.}
    \label{fig:rnfl_seg}
\end{figure*}

\subsubsection{Agreement in choroid area and thickness.} GPET showed very high repeatability for thickness (Pearson $r$=0.9527, MAE=10.4074 $\mu$m) and area (Pearson $r$=0.9726, MAE=0.0486 mm$^2$). DeepGPET achieved slightly lower, yet also very high agreement for both thickness (Pearson $r$=0.8908, MAE=13.3086 $\mu$m) and area (Pearson $r$=0.9082, MAE=0.0699 mm$^2$). \cref{fig:corr_plot} shows correlation plots for thickness and area. DeepGPET's agreement with GPET does not quite reach the repeatability of GPET itself, when used by the same experienced analyst, but it is quite comparable and high in absolute terms. Especially noteworthy is that the MAE for thickness and area is only 21\% lower for thickness and 30\% lower for area for repeated GPET than for DeepGPET Thus, DeepGPET comes quite close to optimal performance, i.e. best case repeatability where the same experienced analyst did both sets of annotation. 

Furthermore, the regression fits in both derived measures for DeepGPET are closer to the identity line than for the repeated GPET measurements. For CT, the linear fit estimated a slope value of 1.043 (95\% confidence interval of 0.895 to 1.192) and intercept of -7.308 $\mu$m (95\% confidence interval of -48.967 $\mu$m to 34.350 $\mu$m). For CA, the linear fit estimated a slope value of 1.01 (95\% confidence interval of 0.878 to 1.137) and an intercept of 0.016 mm$^2$ (95\% confidence interval of -0.195 mm$^2$ to 0.226 mm$^2$). All confidence intervals contain 1 and 0 for the slope and intercepts, respectively, suggesting no systematic bias or proportional difference between GPET and DeepGPET \cite{passing1983new,ranganathan2017common}.

% Response to R2, point 5
\cref{fig:ba_plot} shows the residuals between DeepGPET and the ground truth labels from the held-out test set using Bland-Altman plots \cite{bland1986statistical}. Rahman \cite{rahman2011repeatability} found that intra-rater agreement and inter-rater agreement of subfoveal choroidal thickness measurements were 23$\mu$m and 32$\mu$m, respectively. For CT, only 9.3\% (5 / 54) were greater than 23$\mu$m in absolute value, with 4 of these representing major sources of disagreement. Similarly for CA, the majority of residuals were centred around 0 (mean residual of -0.02mm$^2$), with only 5.5\% (3 / 54) of residuals lying outside the limits of agreement.

\subsection{Qualitative}

\cref{tab:qualitative} shows the results of the adjudication between GPET and DeepGPET. The upper boundary was rated as ``Very good'' for both methods in all 20 cases. However, for the lower boundary, DeepGPET was rated as ``Bad'' in 2 cases for the lower boundary and 1 case for smoothness. Otherwise, both methods performed very similarly. 

\cref{fig:adjudication_examples} shows some examples. In (a), DeepGPET segments more of the temporal region than GPET does, providing a full width segmentation which was preferred by the rater. Additionally, both approaches are able to segment a smooth boundary, even in regions with stroma fluid obscuring the lower boundary (red arrow). 
In (b), the lower boundary for this choroid is very faint and is actually below the majority of the vessels sitting most posterior (red arrow). DeepGPET produced a smooth and concave boundary preferred by the rater, while GPET fell victim to hugging the posterior most vessels in the subfoveal region. 
In (c), DeepGPET rejected the true boundary in the low contrast region (red arrow) and opted for a more well-defined one, while GPET segmented the more uncertain path. Since GPET permits human intervention, there is more opportunity to fine tune it's parameters to fit what the analyst believes is the true boundary. Here, the rater preferred GPET, while DeepGPET's under-confidence led to under-segmentation and to a bad rating.
In (d), the lower boundary is difficult to delineate due to a thick suprachoroidal space (red arrow) and thus a lack of lower boundary definition. Here, the rater gave a bad rating to DeepGPET and preferred GPET, while remarking that GPET actually under-segmented the choroid by intersecting through posterior vessels. 
The choroids in \cref{fig:adjudication_examples}(b--d) are the choroids with the largest CT and CA disagreement between DeepGPET and GPET as observed in \cref{fig:ba_plot}.

\section{Discussion}
We developed DeepGPET, a fully-automatic and efficient method for choroid layer segmentation, by distilling GPET, a clinically validated semi-automatic method. DeepGPET achieved excellent agreement with GPET on held-out data in terms of segmentation and derived choroidal measurements, approaching the repeatability of GPET itself and well within the threshold expected to exceed inter-rater agreement as observed in previous work \cite{rahman2011repeatability}. We also found no significant association between segmentation performance (via Dice score) and choroidal thickness, area and the Heidelberg signal-to-noise quality index in the held-out test set (supplementary \cref{tab:ct_ca_q__tab} and \cref{fig:dice_ct_ca_q}). Most importantly, DeepGPET does not require specialist training and can process images fully-automatically in a fraction of the time, enabling analysis of large scale datasets and potential deployment in clinical practice.

\begin{figure*}[tb]
\centering
  \begin{subfigure}{0.33\textwidth}
  \centering
  \includegraphics[width=\textwidth]{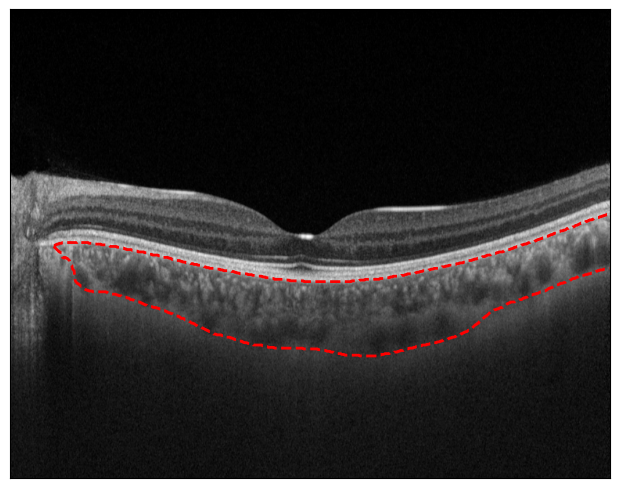}
  \end{subfigure}%
  \begin{subfigure}{0.33\textwidth}
  \centering
  \includegraphics[width=\textwidth]{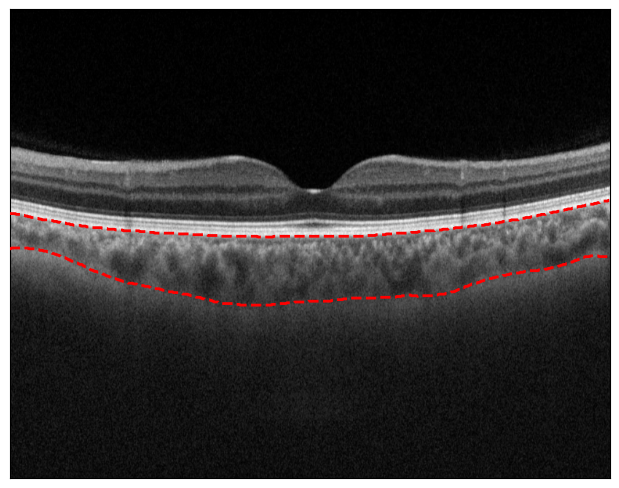}
  \end{subfigure}%
  \begin{subfigure}{0.33\textwidth}
  \centering
  \includegraphics[width=\textwidth]{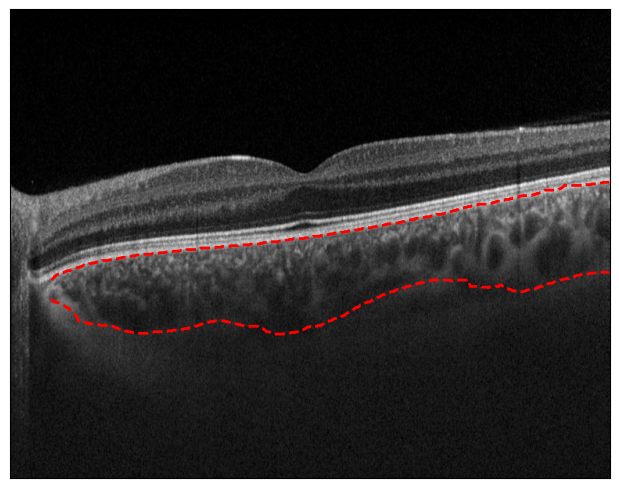}
  \end{subfigure}
\caption{Three OCT B-scan images from a TopCon imaging device, of which two were successful (a--b) and one was not (c).}
\label{fig:topcon_seg}
\end{figure*}

While the observed agreement was very high, it was not perfect. However, even higher agreement with GPET would not necessarily produce a better method as GPET itself is not perfect and even conceptually there is debate around the exact location of choroid-scleral interface (CSI), i.e. the lower choroid boundary in an OCT B-scan. CSI is commonly defined, e.g. by the original authors behind EDI-OCT \citep{spaide2008enhanced}, as the smooth inner boundary between the choroid and sclera, or just below the most posterior vessels but excluding the suprachoroidal space. However, even that definition is still debated and can be hard to discern in practice. Not all choroids are smooth, and there are edge cases like vessels passing from the sclera into the choroid, or stroma fluid obscurations that make the boundary even more ambiguous. These features, coupled with low signal-to-noise ratio and vessel shadowing from superficial retinal vessels, all contribute to the difficult challenge of choroid layer segmentation.

For quantitative analysis of choroidal phenotypes, the specific definition of the CSI is secondary to applying the same, consistent definition across and within patients. Here, fully-automatic methods like DeepGPET provide a large benefit by removing the subjectivity present in semi-automatic methods. Where semi-automatic methods require manual input, two analysts with different understandings of the CSI could produce vastly different segmentations. With DeepGPET, the same image is always segmented in the same way, removing subjectivity.

Initial experiments with other types of OCT imaging have positively indicated DeepGPET's ability to generalise to different visualisations of the choroid. \cref{fig:rnfl_seg} shows a peripapillary scan extracted from the Heidelberg Standard Module, centred on the optic head, with the choroid automatically segmented. \cref{fig:topcon_seg} shows choroid segmentations using DeepGPET for three OCT B-scans from a TopCon device (DRI OCT Triton plus) --- two cases where DeepGPET works well and one case where it does not. This shows some promise in it's usability in scans different to the Heidelberg macular line scans from which it was trained on. We hope in future iterations to extend the training data with scans from different imaging devices and scan locations. We recommend those using DeepGPET on non-Heidelberg images to review the segmentations after ward as a sanity check.

In the present work, we used data from three studies, two OCT devices and included both EDI and non-EDI scans. However, we only used data from subjects that were either healthy or had systemic but not eye disease, to which DeepGPET might not be robust to. In future work, we plan to externally validate DeepGPET and include cases of ocular pathologies. A further limitation is that while GPET has been clinically validated, not all segmentations used for training DeepGPET were entirely perfect. Thus, revisiting some of the existing segmentations and manually improving them to a ``gold standard'' for purposes of training the model could improve DeepGPET. For instance, GPET does not always segment the whole width of the choroid. Interestingly, DeepGPET already is able to do that in some cases (e.g.  \cref{fig:adjudication_examples}(a) and \cref{fig:rnfl_seg}), but also does emulate the incomplete segmentations by GPET in other cases. A model trained on enhanced ``gold standard'' segmentations would produce even better segmentations.

Finally, we have focused on segmentation as it is the most important and most time-consuming step of choroidal analysis. However, the location of the fovea on OCT images needs identified to define the region of interest for derived measurements such as thickness, area and volume. Identifying the fovea is less time-consuming or ambiguous than choroid segmentation, and so we plan to extend DeepGPET to output the fovea location. This would make DeepGPET a fast and efficient end-to-end framework capable of converting a raw OCT image to a set of clinically meaningful segmentation-derived measurements. Likewise, segmenting the choroidal vessels is a very challenging task even for humans and would be prohibitively time-consuming to do manually, but in the future we aim to explore whether DeepGPET can automatically segment the vasculature within the choroid as well.

\section{Conclusion}
Choroid segmentation is a key step in calculating choroidal measurements like thickness and area. Currently, this is commonly done manually which is labour intensive and introduces subjectivity. Semi-automatic methods only partially alleviate both of these problems, and previous fully-automatic methods were not easily accessible for researchers. DeepGPET addresses this gap as a fully-automatic, end-to-end algorithm that does not require manual interventions. DeepGPET provides similar performance as the previously clinically validated, semi-automatic GPET, while being fully-automatic and an order of magnitude faster. This enables the analysis of large scale datasets and potential deployment in clinical practice without necessitating a trained operator. Although the definition of the lower choroid boundary is still subject to debate - especially when it comes to suprachoroidal spaces - the most important consideration is to have a method that consistently applies the same definition across subjects and studies, which DeepGPET as a fully-automatic method provides. As an easily accessible, open-source algorithm for choroid segmentation, DeepGPET will enable researchers to easily calculate choroidal measurements much faster and with less subjectivity than before. 

\section*{Acknowledgements}
J.B. was supported by the Medical Research Council (grant MR/N013166/1) as part of the Doctoral Training Programme in Precision Medicine at the Usher Institute, University of Edinburgh. J.E. was supported by UK Research and Innovation (grant EP/S02431X/1) as part of the Centre of Doctoral Training in Biomedical AI at the School of Informatics, University of Edinburgh. 
For the purpose of open access, the authors have applied a creative commons attribution (CC BY) licence to any author accepted manuscript version arising. The authors would also like to thank all participants in the studies used in this paper, as well as all staff at the Edinburgh Imaging Facility who contributed to image acquisition for this study. We would like to thank Diana Moukaddem, Niall Strang, Lyle Gray (Glasgow Caledonian University) and Paul McGraw (University of Nottingham) for providing the three TopCon OCT B-scan images.

\section{Conflicts of Interest}
The authors declare no conflicts of interest.

\bibliographystyle{unsrt}
\bibliography{references}

\begin{thebibliography}{10}

\bibitem{tan2016state}
Kara-Anne Tan, Preeti Gupta, Aniruddha Agarwal, Jay Chhablani, Ching-Yu Cheng, Pearse~A Keane, and Rupesh Agrawal.
\newblock State of science: choroidal thickness and systemic health.
\newblock {\em Survey of ophthalmology}, 61(5):566--581, 2016.

\bibitem{balmforth2016chorioretinal}
Craig Balmforth, Job~JMH van Bragt, Titia Ruijs, James~R Cameron, Robert Kimmitt, Rebecca Moorhouse, Alicja Czopek, May~Khei Hu, Peter~J Gallacher, James~W Dear, et~al.
\newblock Chorioretinal thinning in chronic kidney disease links to inflammation and endothelial dysfunction.
\newblock {\em JCI insight}, 1(20), 2016.

\bibitem{robbins2021choroidal}
Cason~B Robbins, Dilraj~S Grewal, Atalie~C Thompson, James~H Powers, Srinath Soundararajan, Hui~Yan Koo, Stephen~P Yoon, Bryce~W Polascik, Andy Liu, Rupesh Agrawal, et~al.
\newblock Choroidal structural analysis in alzheimer disease, mild cognitive impairment, and cognitively healthy controls.
\newblock {\em American Journal of Ophthalmology}, 223:359--367, 2021.

\bibitem{eghtedar2022update}
Reza~Alizadeh Eghtedar, Mahdad Esmaeili, Alireza Peyman, Mohammadreza Akhlaghi, and Seyed~Hossein Rasta.
\newblock An update on choroidal layer segmentation methods in optical coherence tomography images: a review.
\newblock {\em Journal of Biomedical Physics \& Engineering}, 12(1):1, 2022.

\bibitem{masood2018automatic}
Saleha Masood, Bin Sheng, Ping Li, Ruimin Shen, Ruogu Fang, and Qiang Wu.
\newblock Automatic choroid layer segmentation using normalized graph cut.
\newblock {\em IET Image Processing}, 12(1):53--59, 2018.

\bibitem{salafian2018automatic}
Bahareh Salafian, Rahele Kafieh, Abdolreza Rashno, Mohsen Pourazizi, and Saeid Sadri.
\newblock Automatic segmentation of choroid layer in edi oct images using graph theory in neutrosophic space.
\newblock {\em arXiv preprint arXiv:1812.01989}, 2018.

\bibitem{kajic2012automated}
Vedran Kaji{\'c}, Marieh Esmaeelpour, Boris Pova{\v{z}}ay, David Marshall, Paul~L Rosin, and Wolfgang Drexler.
\newblock Automated choroidal segmentation of 1060 nm oct in healthy and pathologic eyes using a statistical model.
\newblock {\em Biomedical optics express}, 3(1):86--103, 2012.

\bibitem{wang2017automatic}
Chuang Wang, Ya~Xing Wang, and Yongmin Li.
\newblock Automatic choroidal layer segmentation using markov random field and level set method.
\newblock {\em IEEE journal of biomedical and health informatics}, 21(6):1694--1702, 2017.

\bibitem{srinath2014automated}
Nizampatnam Srinath, A~Patil, V~Kiran Kumar, Soumya Jana, Jay Chhablani, and Ashutosh Richhariya.
\newblock Automated detection of choroid boundary and vessels in optical coherence tomography images.
\newblock In {\em 2014 36th Annual International Conference of the IEEE Engineering in Medicine and Biology Society}, pages 166--169. IEEE, 2014.

\bibitem{george2019two}
Neetha George and CV~Jiji.
\newblock Two stage contour evolution for automatic segmentation of choroid and cornea in oct images.
\newblock {\em Biocybernetics and biomedical Engineering}, 39(3):686--696, 2019.

\bibitem{danesh2014segmentation}
Hajar Danesh, Raheleh Kafieh, Hossein Rabbani, and Fedra Hajizadeh.
\newblock Segmentation of choroidal boundary in enhanced depth imaging octs using a multiresolution texture based modeling in graph cuts.
\newblock {\em Computational and mathematical methods in medicine}, 2014, 2014.

\bibitem{chen2017automated}
Min Chen, Jiancong Wang, Ipek Oguz, Brian~L VanderBeek, and James~C Gee.
\newblock Automated segmentation of the choroid in edi-oct images with retinal pathology using convolution neural networks.
\newblock In {\em Fetal, Infant and Ophthalmic Medical Image Analysis: International Workshop, FIFI 2017, and 4th International Workshop, OMIA 2017, Held in Conjunction with MICCAI 2017, Qu{\'e}bec City, QC, Canada, September 14, Proceedings 4}, pages 177--184. Springer, 2017.

\bibitem{sui2017choroid}
Xiaodan Sui, Yuanjie Zheng, Benzheng Wei, Hongsheng Bi, Jianfeng Wu, Xuemei Pan, Yilong Yin, and Shaoting Zhang.
\newblock Choroid segmentation from optical coherence tomography with graph-edge weights learned from deep convolutional neural networks.
\newblock {\em Neurocomputing}, 237:332--341, 2017.

\bibitem{masood2019automatic}
Saleha Masood, Ruogu Fang, Ping Li, Huating Li, Bin Sheng, Akash Mathavan, Xiangning Wang, Po~Yang, Qiang Wu, Jing Qin, et~al.
\newblock Automatic choroid layer segmentation from optical coherence tomography images using deep learning.
\newblock {\em Scientific reports}, 9(1):3058, 2019.

\bibitem{al2017novel}
Baidaa Al-Bander, Bryan~M Williams, Majid~A Al-Taee, Waleed Al-Nuaimy, and Yalin Zheng.
\newblock A novel choroid segmentation method for retinal diagnosis using deep learning.
\newblock In {\em 2017 10th International Conference on Developments in eSystems Engineering (DeSE)}, pages 182--187. IEEE, 2017.

\bibitem{chen2022application}
Hung-Ju Chen, Yu-Len Huang, Siu-Lun Tse, Wei-Ping Hsia, Chung-Hao Hsiao, Yang Wang, and Chia-Jen Chang.
\newblock Application of artificial intelligence and deep learning for choroid segmentation in myopia.
\newblock {\em Translational Vision Science \& Technology}, 11(2):38--38, 2022.

\bibitem{zheng2021deep}
Gu~Zheng, Yanfeng Jiang, Ce~Shi, Hanpei Miao, Xiangle Yu, Yiyi Wang, Sisi Chen, Zhiyang Lin, Weicheng Wang, Fan Lu, et~al.
\newblock Deep learning algorithms to segment and quantify the choroidal thickness and vasculature in swept-source optical coherence tomography images.
\newblock {\em Journal of Innovative Optical Health Sciences}, 14(01):2140002, 2021.

\bibitem{engelmann2022robust}
Justin Engelmann, Ana Villaplana-Velasco, Amos Storkey, and Miguel~O Bernabeu.
\newblock Robust and efficient computation of retinal fractal dimension through deep approximation.
\newblock In {\em Ophthalmic Medical Image Analysis: 9th International Workshop, OMIA 2022, Held in Conjunction with MICCAI 2022, Singapore, Singapore, September 22, 2022, Proceedings}, pages 84--93. Springer, 2022.

\bibitem{burke2021edge}
Jamie Burke and Stuart King.
\newblock Edge tracing using gaussian process regression.
\newblock {\em IEEE Transactions on Image Processing}, 31:138--148, 2021.

\bibitem{burke2023evaluation}
Jamie Burke, Dan Pugh, Tariq Farrah, Charlene Hamid, Emily Godden, Tom MacGillivray, Neeraj Dhaun, Kenneth Baillie, Stuart King, and Ian J.~C. MacCormick.
\newblock Evaluation of an automated choroid segmentation algorithm in a longitudinal kidney donor and recipient cohort, 2023.

\bibitem{ss2019octtools}
Sarah Patterson.
\newblock Oct-tools.
\newblock \url{https://github.com/sarastokes/OCT-tools}, 2019.

\bibitem{brandt2018octmarker}
Alexander Brandt.
\newblock Oct-marker.
\newblock \url{https://github.com/neurodial/OCT-Marker/tree/master}, 2018.

\bibitem{kugelman2019automatic}
Jason Kugelman, David Alonso-Caneiro, Scott~A Read, Jared Hamwood, Stephen~J Vincent, Fred~K Chen, and Michael~J Collins.
\newblock Automatic choroidal segmentation in oct images using supervised deep learning methods.
\newblock {\em Scientific reports}, 9(1):13298, 2019.

\bibitem{xuan2023deep}
Meng Xuan, Wei Wang, Danli Shi, James Tong, Zhuoting Zhu, Yu~Jiang, Zongyuan Ge, Jian Zhang, Gabriella Bulloch, Guankai Peng, et~al.
\newblock A deep learning--based fully automated program for choroidal structure analysis within the region of interest in myopic children.
\newblock {\em Translational Vision Science \& Technology}, 12(3):22--22, 2023.

\bibitem{mazzaferri2017open}
Javier Mazzaferri, Luke Beaton, Gis{\`e}le Hounye, Diane~N Sayah, and Santiago Costantino.
\newblock Open-source algorithm for automatic choroid segmentation of oct volume reconstructions.
\newblock {\em Scientific reports}, 7(1):42112, 2017.

\bibitem{dhaun2014optical}
Neeraj Dhaun.
\newblock Optical coherence tomography and nephropathy: The octane study.
\newblock \url{https://clinicaltrials.gov/ct2/show/NCT02132741}, 2014.
\newblock ClinicalTrials.gov identifier: NCT02132741. Updated November 4, 2022. Accessed May 31st, 2023.

\bibitem{pearson2022multi}
Thomas Pearson, Yingdi Chen, Baljean Dhillon, Siddharthan Chandran, Jano van Hemert, and Tom MacGillivray.
\newblock Multi-modal retinal scanning to measure retinal thickness and peripheral blood vessels in multiple sclerosis.
\newblock {\em Scientific Reports}, 12(1):20472, 2022.

\bibitem{ronneberger2015u}
Olaf Ronneberger, Philipp Fischer, and Thomas Brox.
\newblock U-net: Convolutional networks for biomedical image segmentation.
\newblock In {\em Medical Image Computing and Computer-Assisted Intervention--MICCAI 2015: 18th International Conference, Munich, Germany, October 5-9, 2015, Proceedings, Part III 18}, pages 234--241. Springer, 2015.

\bibitem{howard2019searching}
Andrew Howard, Mark Sandler, Grace Chu, Liang-Chieh Chen, Bo~Chen, Mingxing Tan, Weijun Wang, Yukun Zhu, Ruoming Pang, Vijay Vasudevan, et~al.
\newblock Searching for mobilenetv3.
\newblock In {\em Proceedings of the IEEE/CVF international conference on computer vision}, pages 1314--1324, 2019.

\bibitem{loshchilov2017decoupled}
Ilya Loshchilov and Frank Hutter.
\newblock Decoupled weight decay regularization.
\newblock {\em arXiv preprint arXiv:1711.05101}, 2017.

\bibitem{Iakubovskii2019}
Pavel Iakubovskii.
\newblock Segmentation models pytorch.
\newblock \url{https://github.com/qubvel/segmentation\_models.pytorch}, 2019.

\bibitem{rw2019timm}
Ross Wightman.
\newblock Pytorch image models.
\newblock \url{https://github.com/rwightman/pytorch-image-models}, 2019.

\bibitem{early1991early}
Early Treatment Diabetic Retinopathy Study~Research Group et~al.
\newblock Early treatment diabetic retinopathy study design and baseline patient characteristics: Etdrs report number 7.
\newblock {\em Ophthalmology}, 98(5):741--756, 1991.

\bibitem{passing1983new}
H~Passing and W~Bablok.
\newblock A new biometrical procedure for testing the equality of measurements from two different analytical methods. application of linear regression procedures for method comparison studies in clinical chemistry, part i.
\newblock 1983.

\bibitem{ranganathan2017common}
Priya Ranganathan, CS~Pramesh, and Rakesh Aggarwal.
\newblock Common pitfalls in statistical analysis: Measures of agreement.
\newblock {\em Perspectives in clinical research}, 8(4):187, 2017.

\bibitem{bland1986statistical}
J~Martin Bland and DouglasG Altman.
\newblock Statistical methods for assessing agreement between two methods of clinical measurement.
\newblock {\em The lancet}, 327(8476):307--310, 1986.

\bibitem{rahman2011repeatability}
Waheeda Rahman, Fred~Kuanfu Chen, Jonathan Yeoh, Praveen Patel, Adnan Tufail, and Lyndon Da~Cruz.
\newblock Repeatability of manual subfoveal choroidal thickness measurements in healthy subjects using the technique of enhanced depth imaging optical coherence tomography.
\newblock {\em Investigative ophthalmology \& visual science}, 52(5):2267--2271, 2011.

\bibitem{spaide2008enhanced}
Richard~F Spaide, Hideki Koizumi, and Maria~C Pozonni.
\newblock Enhanced depth imaging spectral-domain optical coherence tomography.
\newblock {\em American journal of ophthalmology}, 146(4):496--500, 2008.

\end{thebibliography}

\onecolumn
 
\setcounter{figure}{0}
\renewcommand{\thefigure}{S\arabic{figure}}
\setcounter{table}{0}
\renewcommand{\thetable}{S\arabic{table}}
\pagebreak

\section*{Supplementary Material}

\begin{figure}[H]
\centering
  \begin{subfigure}{0.37\textwidth}
  \centering
  \includegraphics[width=\textwidth]{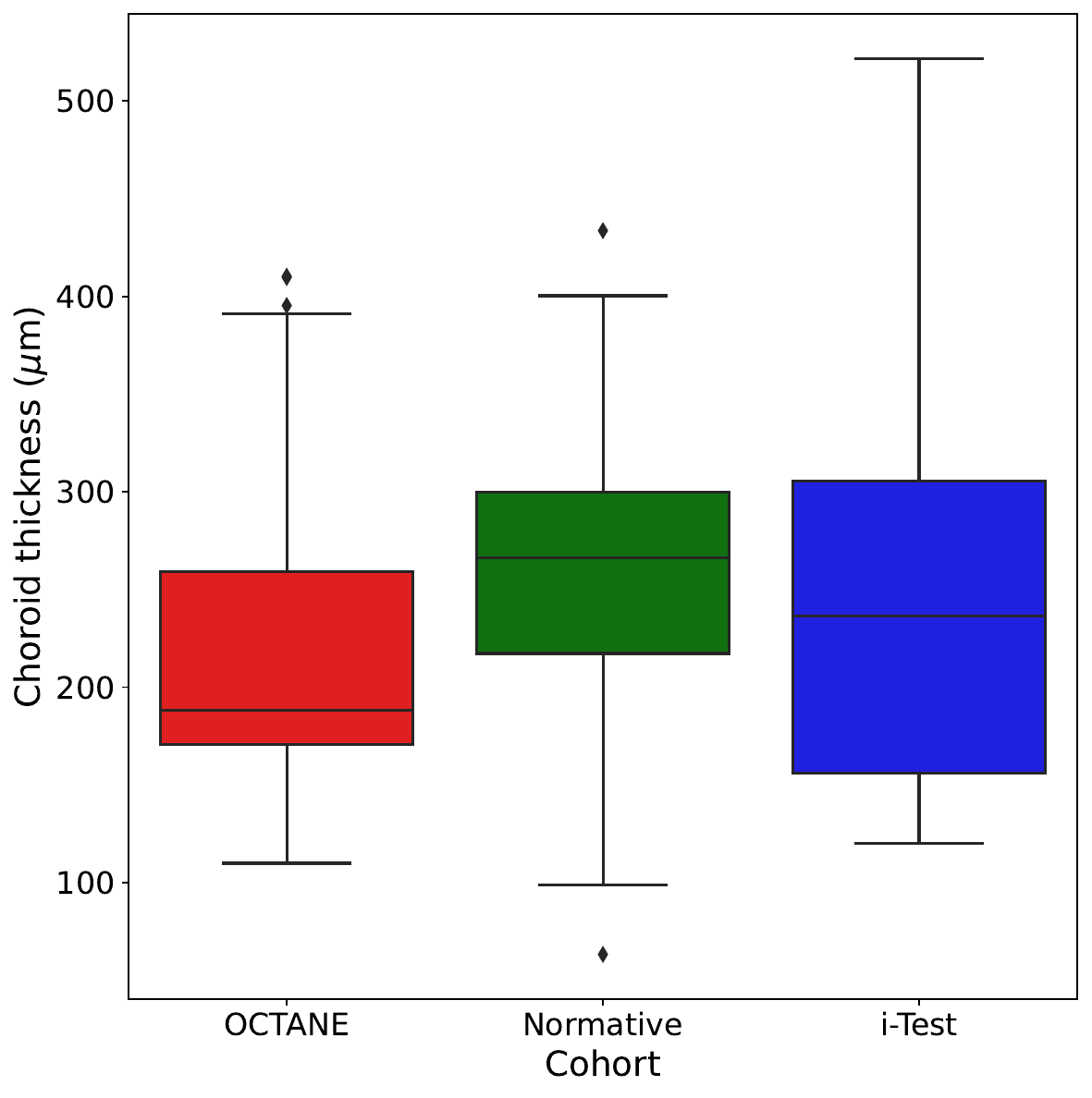}
  \end{subfigure}%
  \begin{subfigure}{0.37\textwidth}
  \centering
  \includegraphics[width=\textwidth]{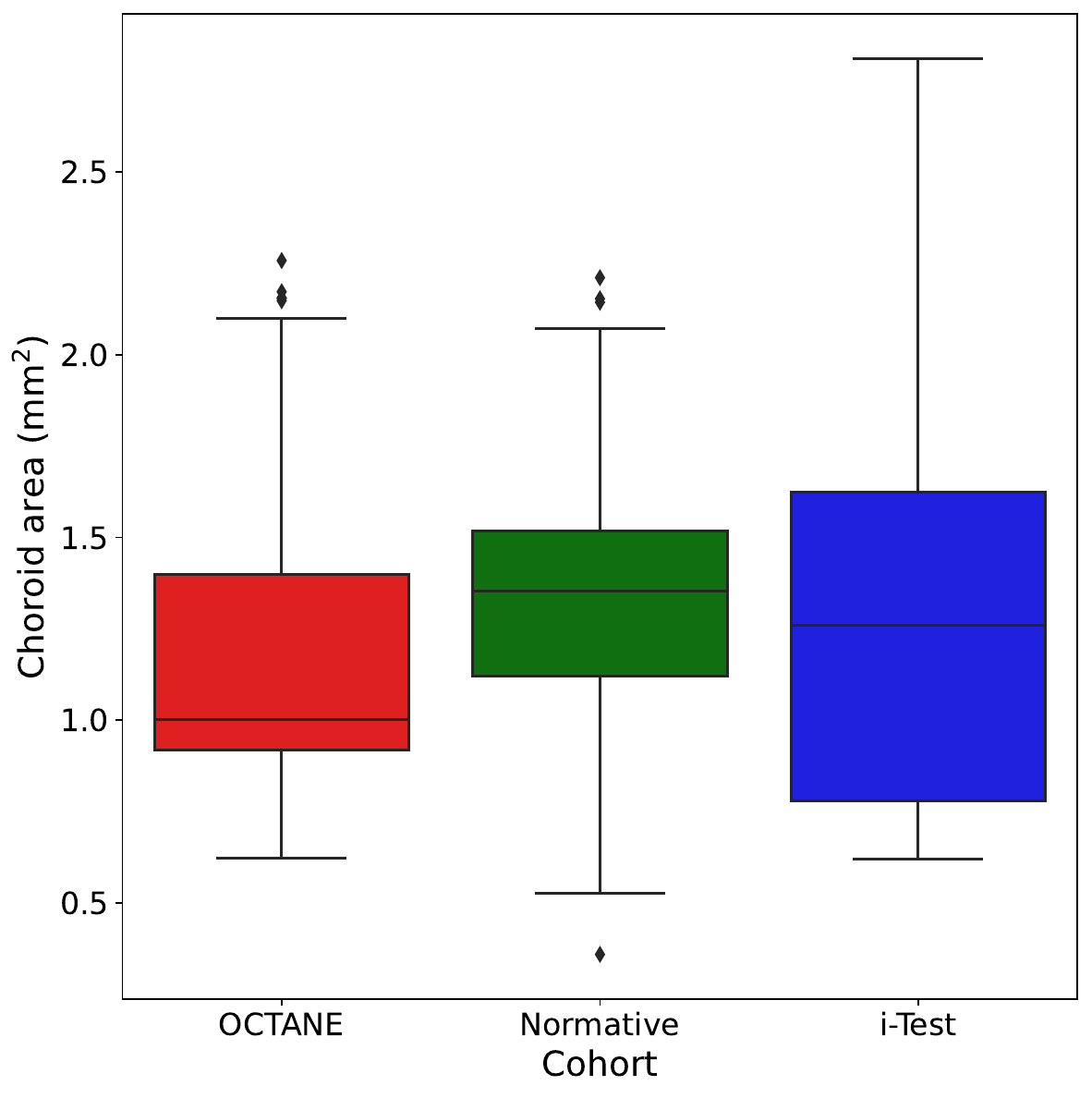}
  \end{subfigure}
\caption{Box-plot distribution plots of choroid thickness (left) and area (right) of the three datasets, OCTANE, i-Test and Normative.}
\label{fig:bp_chor_charac}
\end{figure}

\begin{table}[H]
\centering
{\small
\begin{tabular}{@{}lccc@{}}
\toprule
\hspace{4em}                    & \hspace{1em}OCTANE\hspace{1.3em}       & \hspace{1em}i-Test\hspace{1.3em}     &\hspace{1em}Normative\hspace{1.3em} \\ \midrule
Choroid thickness ($\mu$m)  & 220.4 (68.6)$^\text{**,***}$  & 245.3 (89.4)$^\text{**}$ &  256.8 (65.9)$^\text{***}$ \\
Choroid area (mm$^2$)  & 1.19 (0.37)$^\text{*,**}$   & 1.30 (0.50)$^\text{*}$  & 1.32 (0.35)$^\text{**}$    \\
\bottomrule
\end{tabular}%
}
\caption{Choroidal characteristics of the three datasets used to build DeepGPET. Values are shown as mean with standard deviation values in parenthesis. Statistically significant results were computed using the T-test comparing the means of independent samples; * : \textit{P}$<0.01$, ** : \textit{P}$<0.001$, *** : \textit{P}$<0.0001$.}
\label{tab:chor_characs}
\end{table}

\begin{table}[H]
\centering
{\small
\begin{tabular}{@{}lccc|c@{}}
\toprule
\hspace{4em}                    & \hspace{1em}Training\hspace{1.3em}       & \hspace{1em}Validation\hspace{1.3em}     &\hspace{1em}Testing\hspace{1.3em}  & Total        \\ \midrule
\multicolumn{1}{l}{Subjects}            & 66           & 9          & 7         & 82           \\
\multicolumn{1}{r}{Male/Female}         & 32 / 34     & 7 / 2      & 5 / 2    & 44 / 38     \\
\multicolumn{1}{r}{Right/Left eyes}     & 66 / 27       & 9 / 4      & 6 / 3    & 81 / 34      \\
\multicolumn{1}{r}{Standard/FLEX Device} & 63 / 3 & 8 / 1 & 6 / 1 & 77 / 5 \\
\multicolumn{1}{r}{EDI/non-EDI} & 51 / 15 & 6 / 3 & 5 / 2 & 62 / 20 \\
\multicolumn{1}{r}{Age (mean (SD))}     & 49.2 (11.4) & 40.2 (8.2) & 44.9 (8.2) & 48.0 (11.2) \\
\multicolumn{1}{r}{OCTANE cohort} & 38 & 5 & 4 & 47 \\
\multicolumn{1}{r}{i-Test cohort} & 3 & 1 & 1 & 5 \\
\multicolumn{1}{r}{Normative cohort} & 25 & 3 & 2 & 30 \\
Horizontal/Vertical scans & 202 / 57      & 19 / 8    & 18 / 5    & 239 / 70     \\
Volume scans           & 344          & 31        & 31         & 406          \\
Total B-scans         & 603          & 58        & 54        & 715          \\ 
\bottomrule
\end{tabular}%
}
\caption{Overview of population and image characteristics of the training, validation and test sets.}
\label{supp_tab:tvt_demo_tab}
\end{table}

\begin{table}[H]
\centering{\small
\begin{tabular}{@{}lllll@{}}
\toprule
  & \multicolumn{2}{c}{Pearson} & \multicolumn{2}{c}{Spearman} \\
\cmidrule(l){2-3}\cmidrule(l){4-5}
& $r_p$ & \textit{P}-value & $r_s$ & \textit{P}-value \\
\midrule
Choroid thickness  & 0.22 & 0.11 & 0.08 & 0.57\\
Choroid area  & 0.23 & 0.09 & 0.12 & 0.37\\
Quality  & -0.11 & 0.43 & 0.001 & 0.99\\
\bottomrule
\end{tabular}%
}
\caption{Pearson and Spearman correlation coefficients between Dice scores against Heidelberg-measured quality index, choroid thickness and area in the held-out test set between DeepGPET and ground truths generated from GPET.}
\label{tab:ct_ca_q__tab}
\end{table}

\begin{figure}[H]
    \centering
    \includegraphics[width=\textwidth]{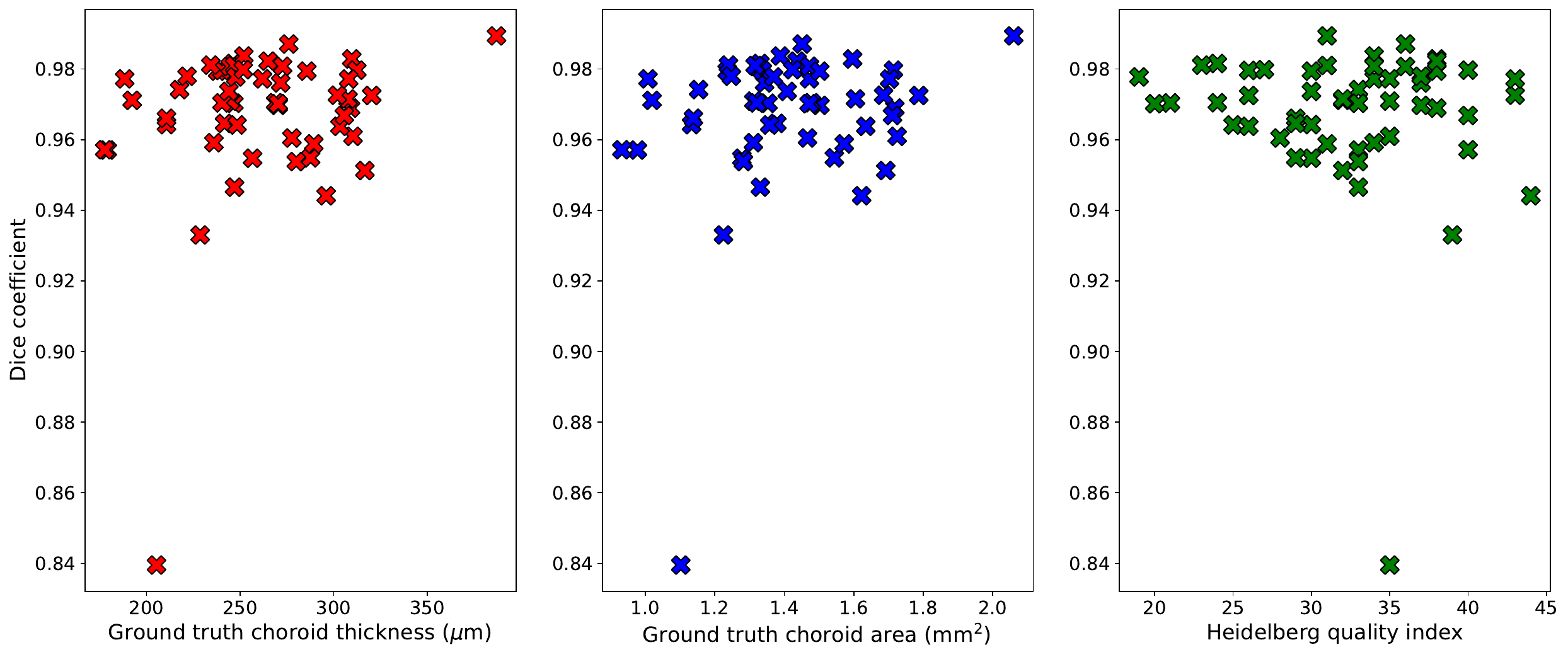}
    \caption{Test set Dice scores plotted against choroid thickness (left), area (middle) and Heidelberg-measured quality index (right) in the held-out test set. The outlier Dice score of approximately 0.84, is the dice score between DeepGPET and GPET from figure 4(d).}
    \label{fig:dice_ct_ca_q}
\end{figure}
\end{document}